\documentclass[journal=jacsat,manuscript=article]{achemso}
\usepackage{achemso} 
\setkeys{acs}{super=false}

\usepackage{color,soul}
\usepackage{amsmath}

\author{Ignacio Chi-Dur\'an}
\affiliation[University of Chicago]
{Department of Chemistry, The University of Chicago, Chicago, IL 60637, USA}

\author{Evan J. Villafranca}
\affiliation[University of Chicago]
{Pritzker School of Molecular Engineering, The University of Chicago, Chicago, IL 60637, USA}

\author{David Dang}
\affiliation[University of Chicago]
{Pritzker School of Molecular Engineering, The University of Chicago, Chicago, IL 60637, USA}

\author{Rachelle Rosiles}
\affiliation[University of Chicago]
{Pritzker School of Molecular Engineering, The University of Chicago, Chicago, IL 60637, USA}

\author{Chun Tung Cheung}
\affiliation[University of Chicago]
{Department of Physics, The University of Chicago, Chicago, IL 60637, USA}

\author{Zhiran Zhang}
\affiliation[University of Chicago]
{Pritzker School of Molecular Engineering, The University of Chicago, Chicago, IL 60637, USA}

\author{Jason P. Cleveland}
\affiliation[QDTI]
{Quantum Diamond Technologies Inc., 1466 Main St, Waltham, MA 02451}

\author{Peter C. Maurer}
\email{pmaurer@uchicago.edu}

\affiliation[University of Chicago]
{Pritzker School of Molecular Engineering, The University of Chicago, Chicago, IL 60637, USA}
\alsoaffiliation[ANL]
{Materials Science Division, Argonne National Laboratory, Lemont, IL, 60439, USA}
\alsoaffiliation[CZB]
{CZ Biohub Chicago, LLC, Chicago, IL 60642, USA}

\title{Quantum biosensing on a multiplexed functionalized diamond microarray}

\begin{document}

\begin{abstract}

Quantum sensing with nitrogen-vacancy (NV) centers in diamond promises to revolutionize biological research and medical diagnostics. Thanks to their high sensitivity, NV sensors could, in principle, detect specific binding events with metabolites and proteins in a massively parallel and label-free way, avoiding the complexity of mass spectrometry. Realizing this vision has been hindered by the lack of quantum sensor arrays that unite high-density spatial multiplexing with uncompromising biochemical specificity. Here, we introduce a scalable quantum biosensing platform that overcomes these barriers by integrating the first multiplexed DNA microarray directly onto a subnanometer antifouling diamond surface. The 7×7 DNA array, patterned onto a diamond chip, enables simultaneous detection of 49 distinct biomolecular features with high spatial resolution and reproducibility, as verified by fluorescence microscopy. Molecular recognition is converted into a quantum signal via a target-induced displacement mechanism in which hybridization removes a Gd$^{3+}$-tagged DNA strand, restoring NV center spin relaxation times (T$_1$) and producing a binary quantum readout. This platform establishes a new paradigm for high-throughput, multiplexed quantum biosensing and opens the door to advanced molecular diagnostics and large-scale quantum sensor networks operable in complex biological environments. 

\end{abstract}

\section{Introduction}

Quantum sensing with nitrogen-vacancy (NV) centers in diamond offers exceptional sensitivity and nanoscale spatial resolution for detecting magnetic fields \cite{Maze2008-zj, Balasubramanian2008, Staudacher2013, Loretz2014, Rovny2022-al}, electric fields \cite{Dolde2011-vc, Bian2021, Qiu2022-ro}, and temperature under ambient conditions \cite{Kucsko2013-fj, Neumann2013, Toyli2013}. These capabilities have enabled growing applications in biology\cite{Miller2020, Kayci2021, Li2022-gy,Rampersaud2024-we,Cleveland2019-tn}, chemistry\cite{Liu2022-xo, Freire-Moschovitis2023}, and materials science\cite{Ariyaratne2018, Huang2023-zu}. A critical barrier to broader deployment, however, lies in translating NV-based sensing into scalable platforms that can detect multiple biomolecular targets with high specificity and minimal device footprint, all while operating reliably in complex biochemical environments. This requires a finely engineered approach to position highly-multiplexed arrays of target molecules within the sensing radius of an NV center without sacrificing chemical stability or biofunctionality.

Current approaches to biointerface fabrication introduce significant limitations. Conventional protocols often rely on atomic layer deposition (ALD) of oxide films or directly controlling the diamond surface chemistry, both of which are multistep and time-intensive\cite{Xie2022-wd,Rodgers2024-il}. Achieving robust immobilization of biomolecules on diamond with these methods comes at the cost of increased NV–analyte separation or potentially, reduced surface stability. To address these challenges, we developed a single-step silanization protocol that covalently attaches biotin–PEG–silane directly to superficial hydroxyl groups on oxygen-terminated (100)-oriented diamond. This rapid (15-minute) process forms a subnanometer antifouling PEG monolayer with terminal biotin groups, eliminating the need for complex surface processing while enabling site-specific and stable biomolecular binding. 

We leverage this subnanometer functionalization layer to demonstrate two critical advances in NV-based biosensing. First, we pattern a 7 x 7 DNA microarray onto a 2 x 2 mm$^2$ diamond chip, enabling multiplexed detection of 49 distinct single-stranded DNA (ssDNA) targets. Each ssDNA spot functions as an isolated sensing region, allowing selective hybridization with complementary DNA (cDNA) strands. Second, we experimentally demonstrate a new transduction mechanism based on displacement of Gd$^{3+}$-DOTA-labeled reporter strands from the diamond surface. When an unlabeled cDNA target strand binds, it displaces the Gd$^{3+}$-labeled strand from the duplex on the surface, removing a local source of magnetic noise and causing a measurable increase in the NV T$_1$ relaxation time. This label displacement assay transforms molecular recognition events into binary NV spin signals, establishing a direct link between biomolecular target binding and quantum readout. 

\section{Results and Discussion}

\subsection{High density DNA immobilization on diamond surfaces}

\begin{figure}[h]
\includegraphics[width=15cm]{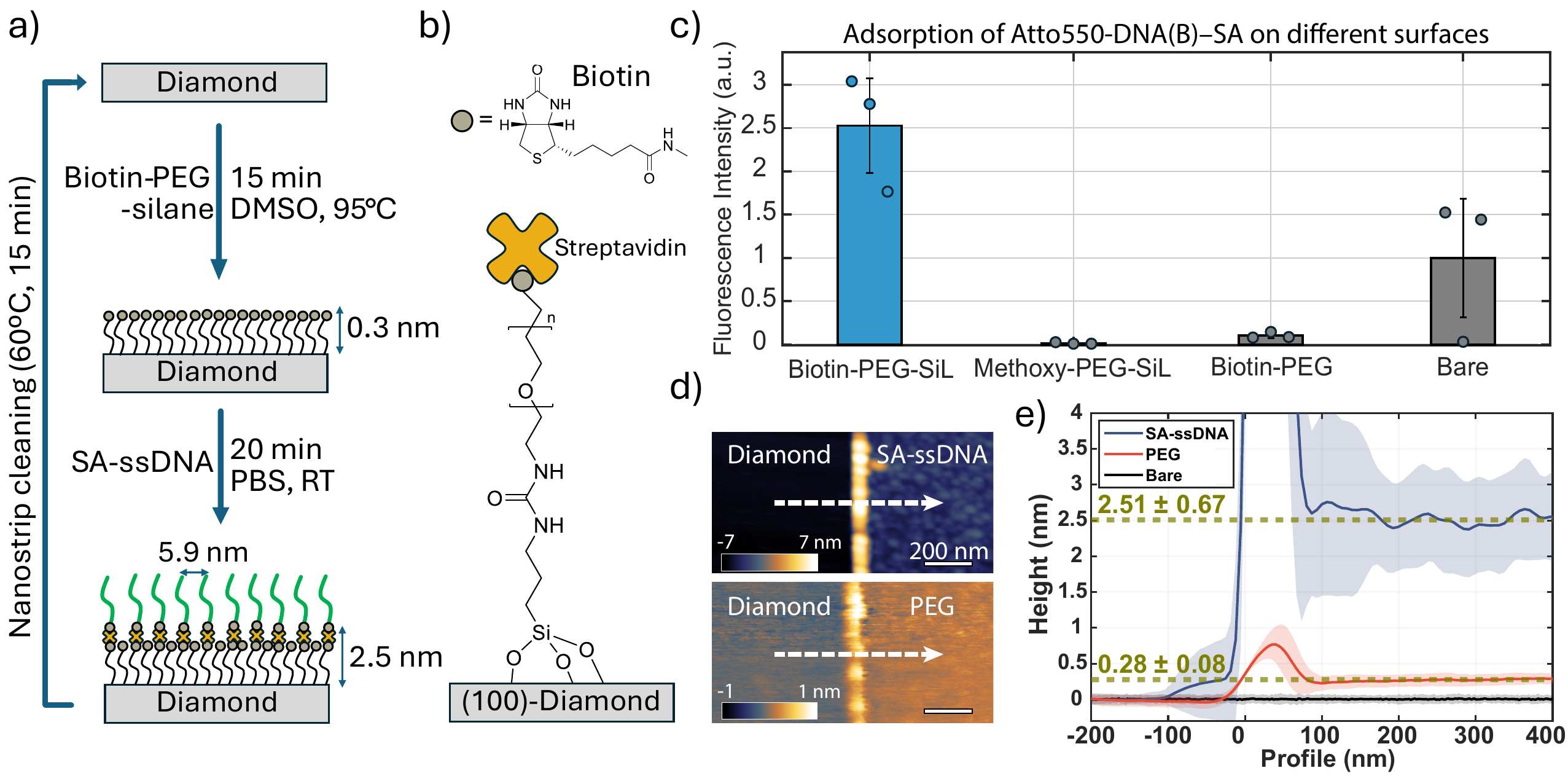}
\centering
\caption{a) Surface functionalization steps to immobilize a high density DNA. b) Chemical structure of biotinylated poly(ethylene glycol) silane covalently attached to the oxygen-terminated diamond surface through native hydroxyl groups. c) Adsorption of DNA(B)–Atto550–streptavidin complex on different surface terminations. Normalized fluorescence intensity is shown for biotin-PEG-silane (specific binding), methoxy-PEG-silane (non-specific binding control), biotin-PEG (no silane control) and bare diamond surface. Error bars represent the standard deviation of the individual measurements (N = 3). d) AFM images after removing a 2x2 $\mu$m$^2$ region using AFM contact mode. e) Height profile of AFM images in d), shading indicates the standard deviation of the cross sections for SA-ssDNA (N=256), PEG (N=512) and bare (N=512) surfaces.}
\label{Figure1}
\end{figure}

We report a new functionalization method that was inspired by Xie \textit{et al.}\cite{Xie2022-wd} and Gidi \textit{et al.}\cite{Gidi2018-be} but eliminates the need for deposition of a nanometer-oxide layer. This step combines silanization and PEGylation into a single reaction by using a biotin-PEG-silane solution in DMSO at 95°C (Figure \ref{Figure1}a). High temperature increases the reactivity of silanes, making it reactive to native hydroxyl groups on the surface of the diamond \cite{Sangtawesin2019-yh, Issa2019-ak} (Figure \ref{Figure1}b). This method reduces the functionalization time from two hours to 15 minutes and thickness by eliminating the metal-oxide intermediate layer as well as the multilayer formation of ethoxysilanes \cite{Sypabekova2022-ra,Zhu2012-pg,Chen2023-ej}. Thus, the spacing between NV centers and the target is minimized, which is favorable for NV-based quantum sensing. X-ray photoelectron spectroscopy confirms successful immobilization and functionalization via the appearance of N 1s and Si 2p peaks at 399.5 eV and 102 eV, respectively, consistent with streptavidin and silane signatures in the literature \cite{Rozyyev2023-in,Williams2013-vd,Meroni2017-pa} (Figure S1). Owing to its organic composition, the layer can be efficiently removed by a brief (15~min) nanostrip treatment, restoring the pristine surface. Furthermore, this method supports the straightforward immobilization of diverse functional groups using commercially available PEG-silanes (e.g., amine, azide, maleimide, alkyne), avoiding lengthy multistep radical-based reactions on diamond surfaces \cite{Rodgers2024-il}. 

Following PEGylation, the diamond surface is able to immobilize a high density of streptavidin, which are conjugated to biotinylated-ssDNA. The strong affinity of the biotinylated PEG surface for streptavidin facilitates the rapid formation of a high-density of ssDNA-conjugated streptavidin complexes within 20 minutes \cite{Green1970-vr,Hirsch2002-rm,Hofmann1984-ff,Hofmann1982-ib}. Atomic force microscope (AFM) scans show a clear characteristic roughness throughout each step (Figure S2), indicating highly homogeneous coverage. Based on scans over a 200 × 200 nm$^2$ area, we estimate 67\% surface coverage of streptavidin, corresponding to an average density of 27,500 ssDNA-conjugated streptavidin molecules per $\mu$m$^2$ (Figure S3). 

In addition, we perform thickness characterization of each layer via atomic force microscopy (AFM) on skive polished diamond surfaces. As shown in Figure \ref{Figure1}d, each coating is removed by scratching a 3 × 3 $\mu$m$^2$ region using the AFM in contact mode, followed by measuring the topography in tapping mode. Figure \ref{Figure1}e shows the height profile with respect to the onset of the edge boundary. Values near zero correspond to the region where the tip scans the bare surface, while positive values indicate the functionalized layer. The biotin-PEG coating exhibits a thickness of 0.28 ± 0.08 nm, consistent with a dried PEG monolayer in a collapsed conformation, spreading laterally across the surface \cite{Hynninen2016-al, Lin1994-dc, Damodaran2012-fd}. Similarly, the ssDNA-conjugated streptavidin layer shows a thickness of 2.5 nm, which is thinner than the reported 4.2 nm thickness of a single streptavidin protein by single-crystal X-ray diffraction \cite{Weber1989-rr, Le-Trong2011-xz}. This reduction is consistent with conformational changes of streptavidin upon dehydration \cite{Li2015-hu}.

Next, we evaluate the biochemical specificity of our functionalized diamond surfaces by probing two critical features: the selective interaction between streptavidin and biotinylated PEG, and the covalent attachment of PEG to the diamond surface via its silane group. To assess streptavidin specificity, we quantify the adsorption of Atto550-labeled ssDNA(B)–streptavidin (Atto550-ssDNA(B)-SA) conjugates on surfaces treated with either biotinylated or non-biotinylated PEG. As shown in Figure \ref{Figure1}c, biotin-PEG-silane enables robust and specific binding of Atto550-ssDNA(B)-SA, while methoxy-PEG-silane, lacking the biotin group, serves as a negative control for non-specific adsorption. Compared to bare diamond, methoxy-PEG-silane reduces fluorescence intensity by a factor of 50, indicating effective suppression of non-specific binding, whereas biotin-PEG-silane enhances the signal by a factor of 2.5, consistent with specific biotin–streptavidin interactions. To verify that PEG was anchored via silane condensation with hydroxyl groups on diamond, we compared Atto550-ssDNA(B)-SA adsorption on surfaces treated with PEG constructs containing or lacking a silane group. Surfaces functionalized with biotin–PEG–silane exhibit a 24-fold higher fluorescence signal than those lacking the silane moiety, confirming grafting through silane chemistry. 

\begin{figure}[H]
\includegraphics[width=9cm]{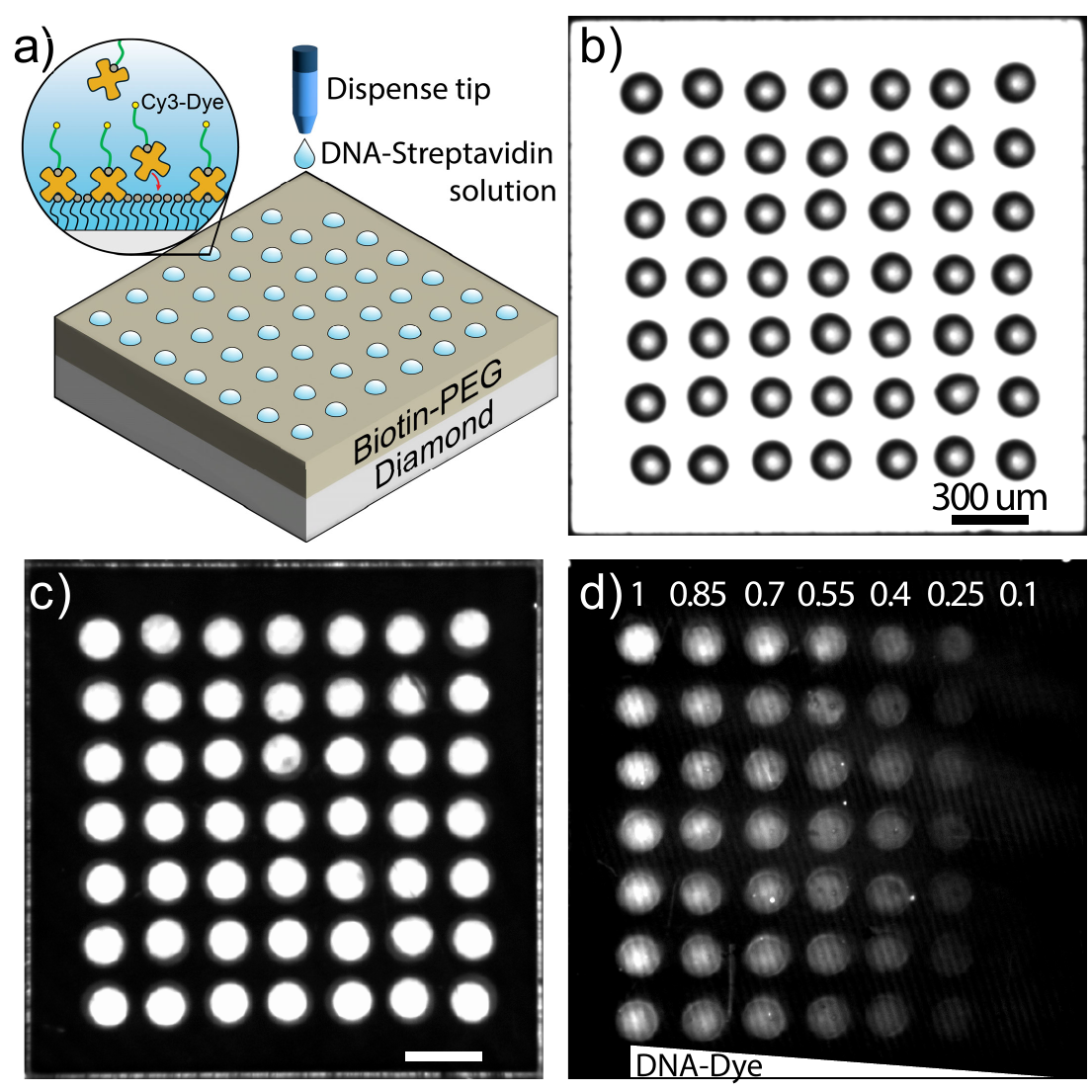}
\centering
\caption{a) Scheme of DNA microarray fabrication using a picoliter dispensing robot. b) 7x7 droplet array on a 2x2 mm$^2$ diamond chip. c) Fluorescent image of Atto550-labeled DNA. d) Binary mixture of Atto550-labeled ssDNA with a non-labeled ssDNA across different rows. Molar fraction of the Atto550-labeled DNA is indicated above each column.}
\label{Figure2}
\end{figure}

The rapid immobilization of ssDNA-conjugated streptavidin enables the fabrication of high-density DNA microarrays on compact diamond surfaces for multiplexed target immobilization. A schematic of the fabrication procedure is depicted in Figure \ref{Figure2}a. A droplet-dispensing robot applies 300-picoliter droplets of DNA-streptavidin solution onto a 2 × 2 mm$^2$ PEGylated diamond surface. We pattern 150 $\mu$m-diameter DNA spots into a tightly packed 7 × 7 array, occupying only ~20\% of the available surface area while generating 49 distinct sensing sites (Figure \ref{Figure2}b). This approach offers exceptional flexibility, permitting the deposition of either single- or multi-component DNA solutions at each spot. As shown in Figure \ref{Figure2}c, Atto550-labeled ssDNA is patterned into highly ordered homogeneous arrays at all sites. Furthermore, multi-component spots are created by co-dispensing Atto550-labeled and unlabeled DNA strands at different molar ratios, enabling tunable fluorescence and customizable spot composition (Figure \ref{Figure2}d).

\begin{figure}[H]
\includegraphics[width=12cm]{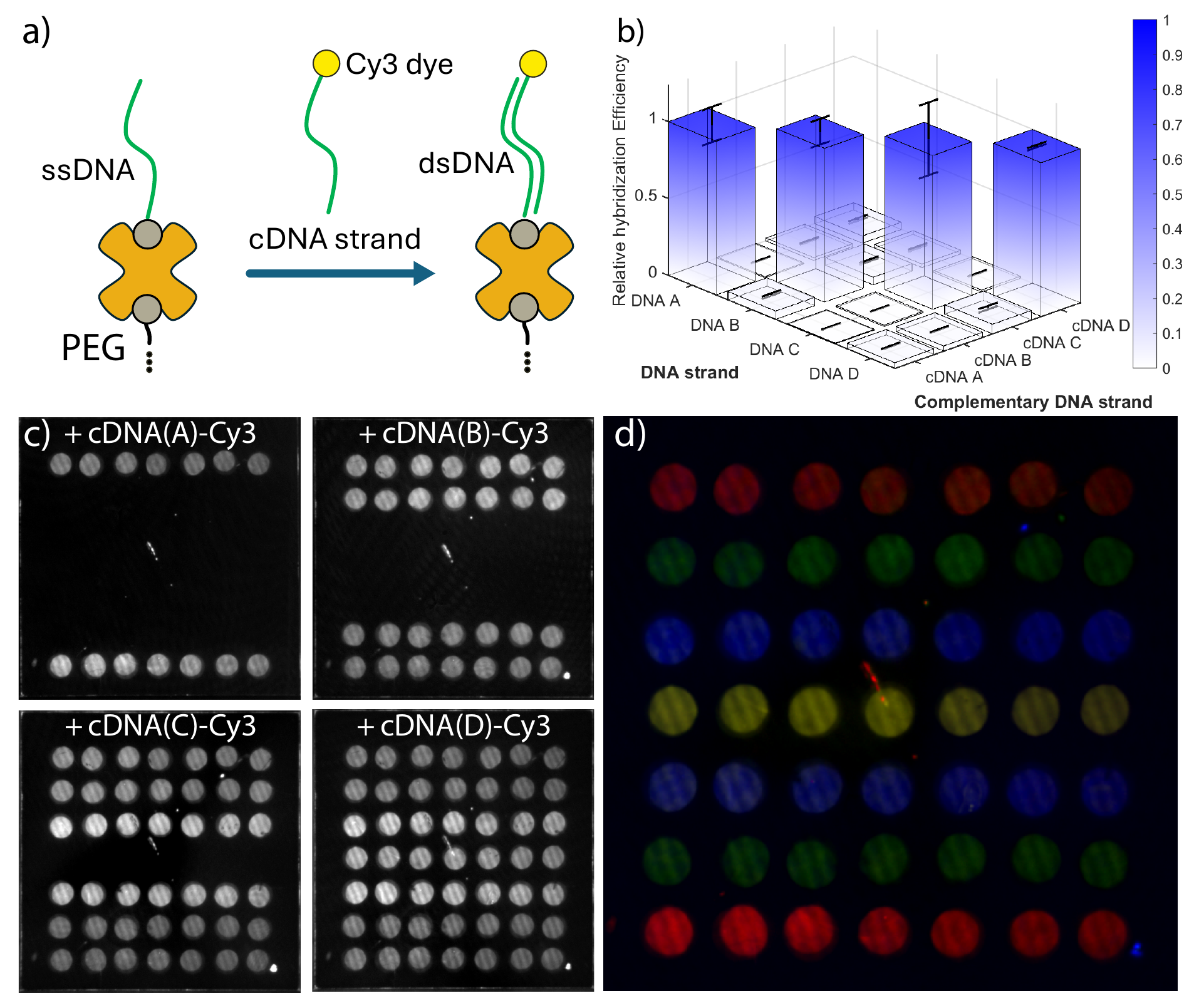}
\centering
\caption{a) Scheme of hybridization of Cy3-labeled cDNA strands on diamond surface. b) Specificity diagram of each DNA strand incubated with complementary DNA strands. Diagonal values indicate high hybridization with its own complementary strand. c) Consecutive hybridization of different Cy3-labeled cDNA strands in the DNA microarray. d) A false-color image of all Cy3-labeled cDNA strands on the diamond surface. The average signal strengths of different cDNAs are calculated from each image in (c). The four resulting images are assigned different colors (red, green, blue, and yellow) and overlaid.
}
\label{Figure3}
\end{figure}

To evaluate the hybridization specificity of our DNA microarray, we symmetrically pattern four spatially distinct regions on the diamond surface, each functionalized with a unique, unlabeled ssDNA sequence (see Table S1). The array is subjected to sequential hybridization steps in which the chip is incubated with Cy3-labeled cDNA strands corresponding to each immobilized sequence (Figure \ref{Figure3}a). After each hybridization step, fluorescence imaging reveals localized signal enhancement exclusively within the designated regions for that target sequence (Figure \ref{Figure3}c), demonstrating the spatial addressability of the array. Quantitative analysis of the fluorescence intensity indicates a hybridization yield of approximately 26\% with respect to direct immobilization of dsDNA (Figure S4). This result is consistent with steric hindrance effects observed in densely grafted oligonucleotide surfaces \cite{Steel1998-cf,Demers2000-eh,Peterson2001-ee}. To assess the extent of non-specific adsorption, the same protocol is repeated with Cy3-labeled non-complementary sequences, and the resulting fluorescence correlation analysis (Figure \ref{Figure3}b) indicates that such interactions contributed less than 6\% of the total signal. These results confirm that our immobilization and hybridization strategy achieves high sequence selectivity with minimal cross-reactivity, providing a robust platform for multiplexed biosensing. To provide an intuitive single-frame summary, the four grayscale images in Figure \ref{Figure3}d are assigned distinct false-colors and overlaid; this composite enables rapid visual verification of specificity.

\begin{figure}[h]
\includegraphics[width=12cm]{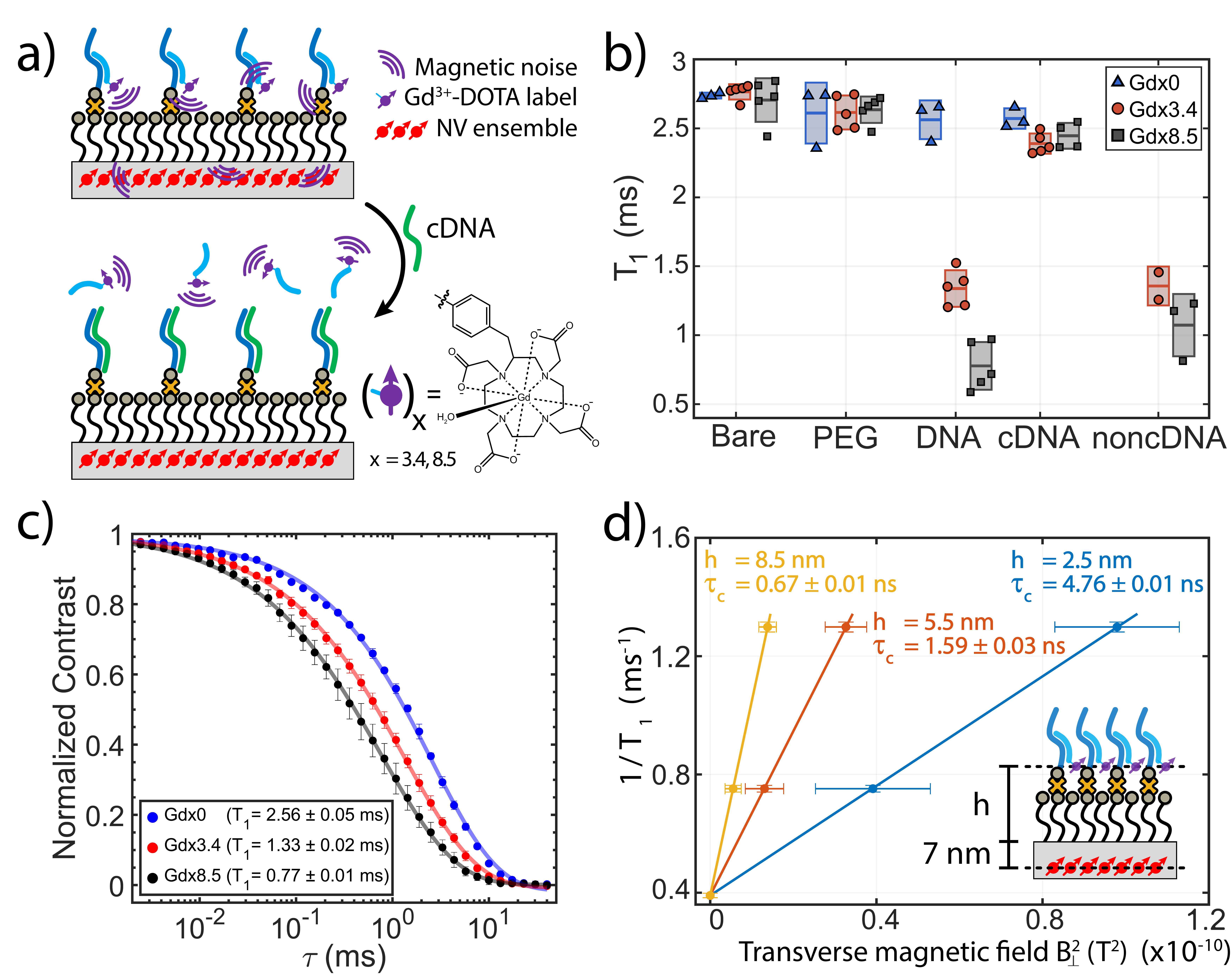}
\centering
\label{Figure4}
\caption{a) Displacement of a short Gd$^{3+}$-DOTA labeled complementary (incumbent) strand by a fully complementary (invader) strand. b) $T_1$ values in each functionalization step. The boxes represent the statistical distribution of each group, where the central horizontal bar denotes the mean \(T_1\) value, and the box height corresponds to the standard deviation. NoncDNA represents a random 24-mer ssDNA as a control. c) $T_1$ fittings for the Gd$^{3+}$-labeled ssDNA in (b). d) Correlation times $\tau_c$ extracted from (c) for three possible NV–Gd$^{3+}$ separations. Horizontal error bars reflect the spread in $B_\perp^2$ arising from the distribution of Gd$^{3+}$ labels in the DNA strands.}
\end{figure}

To establish a generalizable quantum sensing mechanism based on DNA strand displacement, we immobilize a duplex DNA construct in which a long substrate strand is hybridized with a shorter complementary (incumbent) strand as a proof-of-concept (Figure 4a). The incumbent strand is modified with 3.4 ($\pm$1.2) or 8.5 ($\pm$1.3) Gd$^{3+}$-DOTA complexes to modulate the T$_1$ relaxation of near-surface NV centers (see table S2 for details). Sequential T$_1$ measurements confirm minimal changes after PEGylation, whereas immobilization of the Gd$^{3+}$-labeled duplex induces a clear T$_1$ reduction of 47\% (3.4×Gd$^{3+}$) and 70\% (8.5×Gd$^{3+}$) relative to the Gd$^{3+}$-free control (Figure 4b,c). Incubation with a fully complementary “invader” strand displaces the Gd$^{3+}$-labeled strand, restoring 93\% and 95\% of the control T$_1$ values for the 3.4× and 8.5× cases, respectively. No significant recovery is observed when a non-complementary strand is used, confirming the specificity of the displacement mechanism. 

Importantly, this displacement scheme operates without labeling or chemical modification of the target DNA; its presence is transduced into a robust magnetic signal solely through the removal of a proximal Gd$^{3+}$ spin label. Recently proposed for SARS-CoV-2 RNA detection\cite{Li2022-gy}, where viral RNA displaces a Gd$^{3+}$-labeled probe from an NV-tagged nanodiamond surface, and demonstrated with miRNAs and NV-nanodiamonds for cancer detection in Ref. \citenum{Rampersaud2024-we} this strategy exemplifies remarkable versatility. Since the readout depends only on target-driven strand displacement, the approach is inherently generalizable to virtually any biomolecular target using DNA aptamers— including nucleic acids, proteins, and small metabolites— establishing a broadly applicable platform for label-free, multiplexed quantum biosensing on diamond surfaces.

To elucidate the spin dynamics responsible for the observed T$_1$ reduction, we evaluate the variance of the transverse magnetic field ($B_\perp^2$) experienced by the NV center. We model the Gd$^{3+}$ ensemble as an infinite two-dimensional sheet with uniform spin density because the functionalized area is four orders of magnitude larger than the sensing spot, making edge effects negligible (see SI for full derivation). Therefore, the T$_1$ reduction can be calculated based on relations from Refs. (\citenum{Tetienne2013-js,Lu2023-xo}):

\begin{equation}
\frac{1}{T_1} = \frac{1}{T_1^{\text{bulk}}} + 3\gamma_e^2 B_\perp^2 \left( \frac{\tau_c}{1 + \omega_0^2 \tau_c^2} \right)
\end{equation}

\begin{equation}
B_\perp^2 = \left( \frac{\mu_0 \gamma_e \hbar}{4\pi} \right)^2 C_S \frac{2\pi\sigma}{d^4}.
\end{equation}

Here, $\mu_0$ is the vacuum permeability, $\gamma_e$ is the electron gyromagnetic ratio, $C_S = \frac{S(S+1)}{3}$ with $S(\mathrm{Gd}^{3+}) = 7/2$, $\omega_0$ is the NV center transition frequency, and $\tau_c$ is the correlation time of the Gd$^{3+}$ spin bath. The surface Gd$^{3+}$ spin density $\sigma$ and NV–spin separation $d$ were determined from AFM measurements. Because AFM cross-sections are acquired under dry conditions, the measured distances provide a lower bound on the actual NV–Gd$^{3+}$ separation in the experimental environment. We therefore model $d = 7 + h$ nm, where 7 nm is the NV depth and $h$ spans 2.5-8.5 nm to account for the expected increase in spacing under hydrated conditions (Figure 4d). The NV depth of $7 \pm 2$ nm was obtained from implantation–energy correlations \cite{Pezzagna2010-tj}. Fitting $1/T_1$ versus $B_\perp^2$ yields $\tau_c$ values between 0.67 and 4.76 ns, consistent with previous reports for Gd-complex and DNA-bound dyes. Longer $\tau_c$ values (1.59–4.76 ns) are attributed to labels on DNA strands with restricted rotational mobility or local wobbling \cite{Hartmann2014-xj,Sanborn2007-gc,Unruh2005-xz,Gd_correlationtime_Granato2011,Gd_correlationtime_Granato2018,Gd_correlationtime_Li2019}, whereas shorter values ($<$1 ns) are governed by the shorter rotational correlation time of the Gd-complex at room temperature and low fields ($B_0 \sim 0.01$ T) \cite{Rast2001-qb}. Together, these results connect the measured spin dynamics to the underlying biophysics at the nanoscale on our diamond surface, providing a quantitative basis for interpreting NV–spin bath interactions in realistic biomolecular sensing platforms.


\subsection{Conclusion}

We have demonstrated a scalable quantum biosensing platform that integrates high-density DNA microarrays directly onto a subnanometer-functionalized diamond surface, enabling multiplexed molecular recognition and quantum readout. By combining a rapid, direct silanization strategy with site-specific DNA patterning, we achieve robust and reproducible immobilization of 49 distinct sensing elements on a diamond chiplet. Crucially, we establish a target-triggered displacement mechanism in which DNA hybridization removes a paramagnetic Gd$^{3+}$ label, producing a tunable quantum signal by restoring the NV center spin relaxation time. Although we have demonstrated our approach for DNA sensing, our platform can readily be generalized to the detection of a large number of small molecules or proteins. In this case, instead of a complementary DNA strand, a target small molecule or protein would displace a Gd$^{3+}$ label bound to an immobilized DNA aptamer. Our findings address long-standing challenges in NV-based biosensing—namely, surface functionalization, multiplexing, and specificity—and offer a blueprint for high-throughput, label-free molecular detection in complex biological environments. This work paves the way for the development of integrated quantum diagnostics and sensor arrays capable of real-time, parallelized biomolecular analysis, with potential applications ranging from early disease detection and therapeutic drug monitoring to personalized medicine and intraoperative diagnostics.

\section{Methods}

\subsubsection{Diamond functionalization}
Single-crystalline diamonds slabs (2 × 2 × 0.5 mm$^3$, Element Six, electronic grade, Catalog No. 145-500-0385) were sonicated in water for 5 min and cleaned using nanostrip solution at 60°C for 15 minutes. The diamonds were then copiously rinsed with DI water and dried under a nitrogen flux. To remove excess water that remains on the surface of the diamond, the diamonds were soaked in anhydrous acetone (extra dry, ACROS Organics, Catalog No. AC326801000) for at least 5 minutes prior to silanization. The silanization was achieved using freshly prepared 15\% (m/m) solution of Biotin-PEG-Silane (MW 2k, Laysan Bio) in anhydrous DMSO (extra dry, ACROS Organics, Catalog No. 610421000) at 95°C for 15 minutes. Then, the diamonds were thoroughly rinsed with DI water. For ssDNA-SA immobilization, biotinylated ssDNA were mixed with streptavidin to make a 1 $\mu$M solution, in a ratio ssDNA:SA = 1.5 : 1, followed by incubating the diamonds in this solution for 20 minutes and rinsed with DI water. For hybridization experiments, the same procedure was used as described above with an additional culminating incubation process wherein the diamond was immersed in a solution with 0.25 $\mu$M.
The DNA array was done using a non-contact dispensing robot SciTEM (Scienion AG) using the software sciFLEX\_S12h-TEM version 18.06.1. After dispensing the DNA solutions on a biotinylated surface, the diamonds were incubated for one hour. The diamond chips were rinsed thoroughly with DDI water and then observed in a fluorescent microscope in a solution of PBS with 1\% Tween 20. 

\subsubsection{DNA spin labels} 
Gd$^{3+}$-DOTA-labeled ssDNA were made by reacting ssDNA (see complete sequences in the Supplementary information) with Gadolinium (III) S-2-(4-isothiocyanatobenzyl)-1,4,7,10-tetraazacyclododecane-1,4,7,10-tetraacetic acid (Gd-p-SCN-Bn-DOTA) (Macrocyclics, Catalog No. X-207) with 5xNH2- and 10xNH2-terminated ssDNA in a ratio ssDNA:Gd-p-SCN-Bn-DOTA = 1:30 in 100 mM Borate buffer pH 9.2 for 20 hours at room temperature. The Gd$^{3+}$-DOTA-labeled incumbent strands were double-purified using two spin columns (Micro Bio-Spin P-6, Bio-Rad). Degree of labeling was determined by HPLC-MS spectrometry (Novatia LLC, method: LCMS\_20\_min) by quantifying the area of each chromatography peak.

\subsubsection{Surface characterization}
AFM measurements were performed using a Bruker MM8 model. A 3×3 $\mu$m$^2$ square scratch region was created in contact mode using an AppNano AFM probe (f: 200-400 kHz, k: 13-77 N/m, Part number: ACTA-10), with a 1 V deflection setpoint and a resolution of 2000 lines. Subsequently, a 2×2 $\mu$m$^2$ area, centered at the edge of the scratched region, was imaged in tapping mode with a 90° scan angle. The surface termination morphology was characterized using a Cypher ES Atomic Force Microscope (AFM) (Asylum Research, Oxford Instruments). AFM scans were performed over 1 × 1 $\mu$m$^2$ or 0.2 × 0.2 $\mu$m$^2$ areas using AC-240S cantilevers in AC Air Topography mode with the Asylum Research RealTime software (version 19.19.77). Data processing was conducted with Gwyddion software (version 2.66). The streptavidin surface density was calculated based on a protein surface area of 24.36 nm$^2$ \cite{Weber1989-rr,Le-Trong2011-xz}. 

X-ray photoelectron spectroscopy (XPS) measurements were performed on a Thermo Scientific ESCALAB 250 Xi system equipped with an Al K$\alpha$ source. Spectra were acquired using a 200 $\mu$m spot size and a flood gun to mitigate surface charging. High-resolution scans of N 1s, and O 1s regions were collected with 10 accumulations each, while Si 2p spectra were collected with 50 accumulations.

\subsubsection{Optical microscopy}
Fluorescence imaging was performed on a custom-built fluorescence microscope equipped with a 532 nm laser (Coherent Sapphire) and a 1.25× air objective (Olympus AMEP4736) or 60× oil objective (Olympus UPLAPO60XOHR)  in an inverted configuration with epi$-$illumination. Diamond samples were placed inside an 8-well plate (µ-Slide 8 Well, Ibidi USA) with the functionalized side facing up for imaging of the arrays. For experiments in which the oil objective was used (i.e., hybridization correlation matrix and specific and nonspecific binding assays), the diamonds were facing down and imaged through the glass coverslip. For 532-nm excitation (Atto550/Cy3$-$DNA), a ZET532/10x notch filter, a ZT532rdc$-$UF1 dichroic beamsplitter, and an ET575/50m emission filter were used. Images were acquired by an Andor iXon Ultra 888 electron$-$multiplying charge-coupled device (EMCCD) camera (EMCCD cooled down to $-$60°C) with 10 s exposure time and 20 gain. For array experiments using the air objective, the camera was used in 'CCD lowest noise/slow readout' mode with 5 minutes exposure time and same camera temperature. Background subtraction for array experiments were done by removing the diamond and doing imaging under the same imaging conditions and exposure time. For the four color algorithm, the grey-scale images were subtracted from each sequential image to remove signals of other cDNAs after normalization. 

\subsubsection{Quantum sensing experiments}
We used a custom benchtop epifluorescence inverted microscope to perform $T_1$ relaxometry measurements. The NV ensembles are interrogated with a 515 nm laser (Oxxius LBX-515) through an Olympus U Plan Fluorite 100x Oil Immersion Objective (Edmund Optics 86-823) to excite an NV area of approximately 20 $\mu m$ diameter. A dichroic mirror (Chroma T610lpxr) and long-pass filter (Semrock BLP01-594R-25) are used on the collection path to spatially separate and filter out the excitation light from the NV fluorescence, which is collected on an avalanche photodiode (Thorlabs APD410A) and processed using a 500 MHz digitizer (Spectrum Instruments). T$_1$ measurements were done by measuring T$_1$ times in every functionalization step using 10x PBS, maximizing T$_1$ times according to ref. \citenum{Freire-Moschovitis2023-ru}. For SA-ssDNA functionalization steps we use 10x PBS + 1\% Tween 20.

A vector signal generator (Stanford Research Systems SG396) produces the NV microwaves for spin-state control, and an arbitrary waveform generator (Zurich Instruments HDAWG4) is used for IQ modulation of the SG396. A high-power amplifier (Mini-Circuits ZHL-25W-63 +) amplifies the microwaves before reaching the sample, which sits on a custom PCB containing a coplanar waveguide with a loop antenna. A Petri dish is adhered to the waveguide and PCB by PDMS and holds the diamond in PBS solutions for the measurements. A Pulse Streamer 8/2 (Swabian Instruments) controls all experimental pulse sequences.  

Static magnetic fields are created by a pair of permanent neodymium magnets (K\&J Magnetics). One magnet is positioned on either side of the diamond chip, inspired by the design in Bucher et al. (2019). \cite{buchernatprotocols}. The magnets are housed in a custom mount that allows alignment in spherical coordinates. The mount consists of two rotation stages (Thorlabs HDR50) and two linear stages (Zaber LRT0100AL-CT3A). 

\begin{acknowledgement}

We thank Dr. Yasser Gidi for fruitful discussions on surface functionalization and fluorescence imaging, Wojciech Gogacz for the initial characterization of DNA density using $^{32}$P-labeled DNA, and Michelle Hu for early derivations of the dipolar interactions between Gd$^{3+}$ and NV centers. We also thank Stella Wang for microwave waveguide fabrication. 
This work was supported by the National Science Foundation (NSF) under QuBBE QLCI (NSF OMA- 2121044) and the NSF QuSEC program (MPS-2326748 and MPS-2326792). 
This work made use of the shared facilities at the University of Chicago Materials Research Science and Engineering Center, supported by the National Science Foundation under award number DMR-2011854. Also, this work made use of the EPIC and SPID facility (RRID: SCR\_026361) of Northwestern University’s NUANCE Center, which has received support from the SHyNE Resource (NSF ECCS-2025633), the IIN, and Northwestern's MRSEC program (NSF DMR-2308691).

\end{acknowledgement}

\begin{suppinfo}

Supplementary Information provides the complete list of DNA sequences used (Table S1), the distribution of Gd$^{3+}$-DOTA labeling for Incumbent-NH$_2$ constructs (Table S2), high-resolution XPS characterization of functionalized and control diamond surfaces (Figure S1), fluorescence binding assays on different surface terminations (Figure S2), AFM scans of each surface termination of streptavidin coverage (Figure S3), hybridization efficiency measurements (Figure S4), AFM boundary detection (Figure S5), and a full derivation of the dipolar coupling factor for an infinite 2D Gd$^{3+}$ spin sheet.

\end{suppinfo}

\bibliography{acs-achemso.bib}

\end{document}